# Multiscale Roughness of Upper Mantle Discontinuities Inferred from the USArray: Dependence on Tomography Models

Yinzhi Wang[1]*, Gary Pavlis[2]

[1]Texas Advanced Computing Center, The University of Texas at Austin, Austin, TX 78758, USA

[2]Department of Earth & Atmospheric Sciences, Indiana University, Bloomington, IN 47405, USA

## Abstract
We used 3D plane wave migration to image topography of the upper mantle discontinuities at 410 km and 660 km depth that defines the transition zone. We find both discontinuities have topography variation at all resolvable scales. In this paper we examine the dependency of discontinuity roughness on tomography models. We migrated a common USArray data set with a selection of regional, global, surface wave, and body wave tomography models to distinguish different scales of roughness. The objective is to appraise what features are potential artifact from using an inaccurate velocity model. We find that the largest-scale features depend on the choice of tomography model, while smaller-scale features appear to be almost completely independent of current generation models. We suggest that this observation is additional evidence of the existence of small-scale roughness on the upper mantle discontinuities not captured with the data sampling density of the USArray. We find all models based only on body wave travel times alone do not remove a continent scale offset of both discontinuities that correlates with the edge of the craton. We conclude that offset is an artifact linked to underestimation of wave speed in the upper mantle by pure body wave models. Models produced by joint inversion with surface wave dispersion data are less prone to this apparent artifact. The large-scale topography variation is consistent with rigid plate motion models of the subducted Farallon slab underneath North American. Smaller scale topography is found to have larger variation in regions where the vertical mantle flow through the transition zone is implied by transition zone thickness.

## Highlights
- Imaged the mantle transition zone with different tomography models.
- Pure body wave models consistently underestimate large-scale features.
- Small-scale roughness is consistent across models and unlikely to be artifacts.
- Roughness correlates with vertical mantle flow field.

## 1 Introduction
The part of the Earth's mantle now called the transition zone is defined by globally observed discontinuities in seismic wave speed at nominal depths of 410 and 660 km (d410 and d660). Following pioneering early work (Bernal, 1936; Jeffreys, 1952; Ringwood, 1956), it is now generally accepted that d410 is produced from the phase transition of olivine to wadsleyite, and d660 is produced from the phase transition of ringwoodite to an assemblage of bridgmanite and ferropericlase. Simplified textbook diagrams of whole Earth structure can create a misconception that

d410 and d660 are simple spherical shells that bound the transition zone. Improved seismic observations over the past 30 years, however, have shown this is far from true. Soon after modern, digital, broadband data was commonly available a series of global-scale studies revealed large-scale topography variations on both d410 and d660 (Chevrot et al., 1999; Engdahl & Flinn, 1969; Flanagan & Shearer, 1998, 1998; Gu et al., 1998; Revenaugh & Jordan, 1989, 1991; Shearer, 1991, 1993). Later, higher resolution studies with P receiver functions found evidence of extensive depth variation of these discontinuities at much finer scales (Ai et al., 2007; Cao & Levander, 2010; Collier & Helffrich, 1997; Cottaar & Deuss, 2016; Dueker & Sheehan, 1997; Tibi & Wiens, 2005). A common theme in much of the previous work on d410 and d660 is linking discontinuity depths to mantle temperature. It is well established experimentally (e.g. Helffrich, 2000) that the Clapeyron slope for the two phase changes have opposite signs: d410 is predicted to be elevated where temperatures are lower while d660 is predicted to do the opposite with lower temperature.

In the early 1990s another important aspect of the discontinuities surfaced from studies using short-period signals (Benz & Vidale, 1993). Morozova et al. (1999) and Priestley et al. (1994) suggested the d410 and d660 are best defined by a velocity gradient spanning a few tens of kilometers. At the same time others (Benz & Vidale, 1993; Neele, 1996; Vidale et al., 1995; Yamazaki & Hirahara, 1994) suggested the boundaries were sharper boundaries spanning an extent less than 5 km. To reconcile these contradictory observations, it has been suggested the discontinuities could be either a set of random scatterers (Thybo et al., 2003) or a sharp velocity jump superimposed with a gradational transition (Melbourne & Helmberger, 1998). The common denominator is that the upper mantle discontinuities have a fine structure that may or may not be linked to mantle temperature.

Wang & Pavlis (2016) recently added new insight on d410 and d660 through the first application of full 3D migration imaging of both discontinuities. They presented evidence for fine scale roughness on both discontinuities. This paper is, in a sense, a follow-up to that previous paper. The central theme here is to test a hypothesis that the roughness we found earlier is an artifact of an inaccurate velocity model. The approach we used is to compare results from imaging with every suitable tomography model for the US region that was available in digital form. We find that finer scale roughness does not depend on the model strengthening our earlier assertion. The connection between the roughness feature and mantle process suggests such roughness is an indicator of large-scale vertical flow across the transition zone. We also found that there are large differences at scales of 1000 km or more. That observation leads to a secondary contribution of this paper. That is, our results indicate that all current generation tomography models based only on body wave travel times appear to underestimate upper mantle structure. This paper, thus, is important to those trying to understand the fine scale structure of mantle discontinuities from the perspective of mineral physics or geodynamics. Finally, our results provide preliminary hints on how high-resolution results like this can provide new constraints on the mantle flow field under North America.

## 2 Method

We used the same receiver function data set as that in Wang & Pavlis (2016) processed by the same migration algorithm called PWMIG (Pavlis, 2011; Poppeliers & Pavlis, 2003a, 2003b). PWMIG is a full 3D, vector, imaging method. This imaging technique is a variation of the technology of "migration" that is the foundation of all seismic imaging in the oil and gas industry. Details for the imaging workflow are discussed in Wang & Pavlis (2016) and the numerical implementation is described in Pavlis (2011). A commercial seismic reflection interpretation package is used to manually pick the d410 and d660 horizons independently. The version of PWMIG we used for this paper has a different travel time calculation method to improve the precision and performance of the travel time calculations when using 3D velocity models.

The incident P-wave arrival times are stored in what Pavlis (2011) calls a raygrid by extending the GCLgrid (Fan et al., 2006) of pseudostation points downward along a set of ray paths. This approach allows us to use any three-dimensional velocity model. The algorithm uses a form of approximate ray tracing where the ray geometry is computed from a radially symmetric earth model (in our case AK135 (Kennett et al., 1995)). We then compute P and S travel times for imaging as the radially symmetric model time corrected by a perturbation computed as a line integral along the ray path. This is the traditional way of computing the forward problem in body wave tomography extended here to compute lag for P to S conversions. This approximation is reasonable for all the models we used here because all are strongly smoothed by regularization used to stabilize the inversion.

Here, we present images from the same data set migrated with three regional tomography models that cover the entire contiguous US and one global tomography model. The first is the MIT15 model (Burdick et al., 2017), which we used in our previous publication and serves as a reference. MIT15 was generated from only P wave data. P to S converted wave imaging, however, requires both a P and S velocity model. We generated the S wave model for use with MIT15 by assuming a constant Vp/Vs ratio of the perturbations from AK135 of 1.8. The overall Vp/Vs at any point, however, is not constant in this model since Vp/Vs varies with depth in AK135. The second model we used is that estimated by Shen & Ritzwoller (2016) they call US16. It was computed from a joint Bayesian Monte Carlo inversion of Rayleigh wave speeds determined from ambient noise and earthquakes, receiver functions, and Rayleigh wave ellipticity (H/V) measurements. Their model provides good constraints on the crustal and uppermost mantle structure of contiguous US The third regional model is the USSL14 model (Schmandt & Lin, 2014). It is a high-resolution P- and S-wave tomography model of the upper mantle, which uses a surface wave model of the crust and uppermost mantle as a starting model for the inversion. Finally, we used the GyPSuM model (Simmons et al., 2010) as a representative global scale tomography model. It was computed from simultaneous inversion of seismic body wave travel times and geodynamic observations.

We choose to present results with these models among a wider selection of regional and global models for two reasons. 1) Except for MIT15, all these models have both P and S wave velocities derived from the same algorithm, so we can easily use them without worrying about the inconsistency in combining models with different regularization parameters. 2) The set is a good combination of both surface wave and body wave tomography models, as well as regional and global models. All the regional models provide the highest resolution velocity models currently available. We use GyPSuM as a comparison point for global models. All current generation global models are much lower resolution than those generated from USArray data. Furthermore, all have much smaller maximum velocity perturbations than USSL14 and US16, which we show below have the largest effects on d410 and d660. Hence, we show results only from GyPSuM as a representative example.

A final technical addition to our previous work is that we implemented a static correction (Yang, 2016) to further correct the variation of crustal structure using the CRUST1.0 model (Laske et al., 2013) where appropriate. This was done by computing the difference in travel time between the AK135 model and the CRUST1.0 model and put static shift in time to the receiver functions.

# 3    Topography variation on the discontinuities

Figure 1 shows the topography of d410 from the same data migrated with the four different tomography models. In general, all of them show both large scale, which we define here as greater than 500 kilometers, and smaller scale topography variations.

The large-scale features appear to be most dependent on the choice of tomography models. Comparing the results from US16 model (Figure 1B) and MIT15 model (Figure 1A) one sees that the western US resolves to be much shallower in the US16's result, while the eastern US seems to be identical. That difference is clearer in Figure 2A where we show the difference between depths of the two surfaces. The result with USSL14 (Figure 1C) shows more widespread variations than that with MIT15. In USSL14 the d410 in the southwestern US is imaged shallower while in the northeastern US it is imaged deeper. Overall, one can see that d410 is closer to being flat on the large scale with the USSL14 model compared to any of the other results. In addition, Figure 2B shows that with USSL14 there is substantial variation in the topography of d410 at a range of scales. For example, with USSL14 we image d410 at shallower depths than MIT15 around the area near the southwest corner of Wyoming and in a broader area in centered approximately on Kansas City, Missouri. The result from GyPSuM model (Figure 1D), in contrast, looks almost identical to that from MIT15 model. That visual pattern is confirmed quantitatively by Figure 2C. The difference of the two is mostly under 5 km, and there is no conspicuous pattern like that seen in the other difference surfaces (Figure 2C).

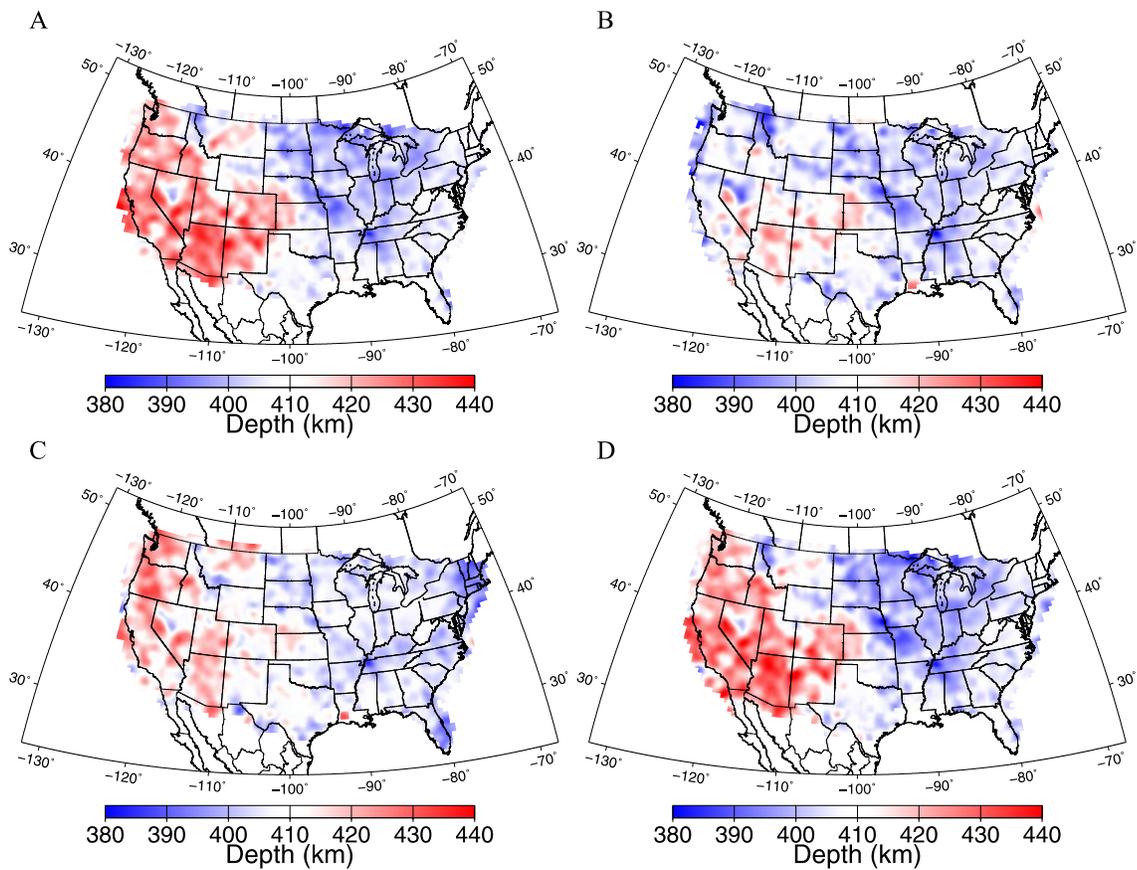

Figure 1. Topography of d410 imaged with four tomography models. The tomography models used in A, B, C, and D are MIT15, US16, USSL14, and GyPSuM respectively. The black lines are state boundaries.

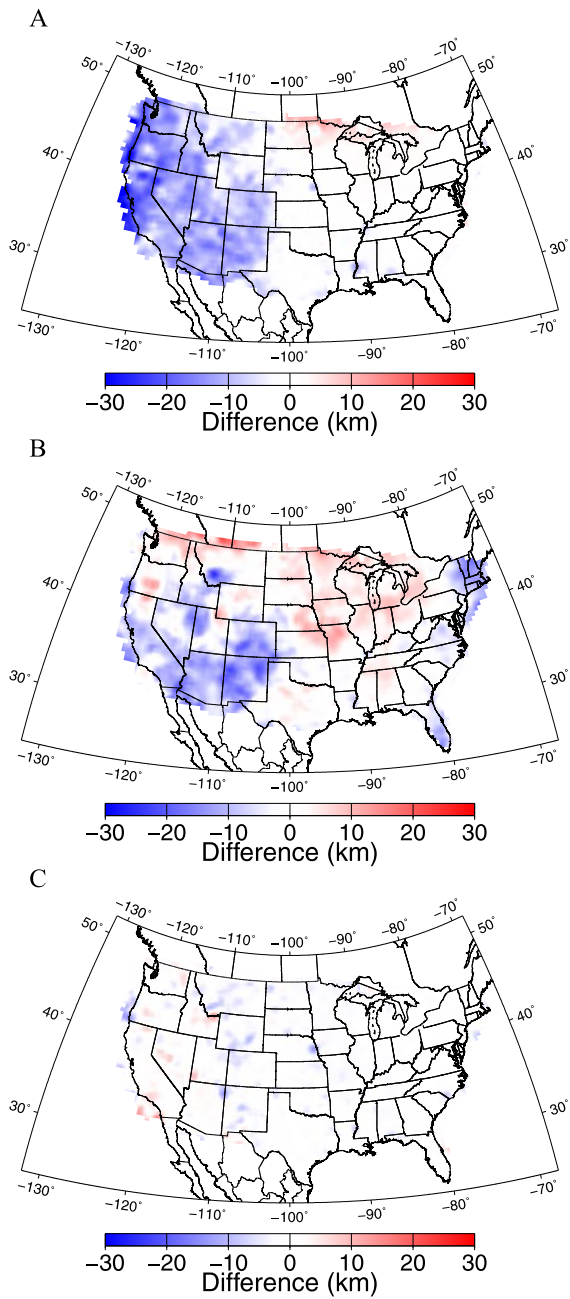

Figure 2. The difference of d410's topography relative to MIT15 reference. These are computed as d410 topography from other models subtract that from MIT15. A is from US16 model. B is from USSL14 model. C is from GyPSuM model.

The same comparisons for d660 are shown in Figure 3 and Figure 4. As for d410 we find most difference between results from MIT15 and US16 are mainly in the west (Figure 4A). On the other hand, there are a lot of smaller scale variations in the eastern US that are not seen in the d410's comparison. The USSL14 model also puts the d660 shallower in the southwestern US and deeper in the northeastern US (Figure 4B). While the large-scale trends mostly follow the patterns seen in the difference map for d410 (Figure 2), there are more variations in depth differences at a tens of kilometers' scale on d660 that have smaller amplitude. The same comparison with the GyPSuM model shows variations mostly within 5 km range (Figure 4C) that are similar to our results for d410. Comparison of Figure 4C to Figure 2C suggests the small features are random and likely related only to variations in manual picks we made to define the two surfaces. On the other hand, there are topography variations around 100 km size in the results from all four models that appear to be consistent.

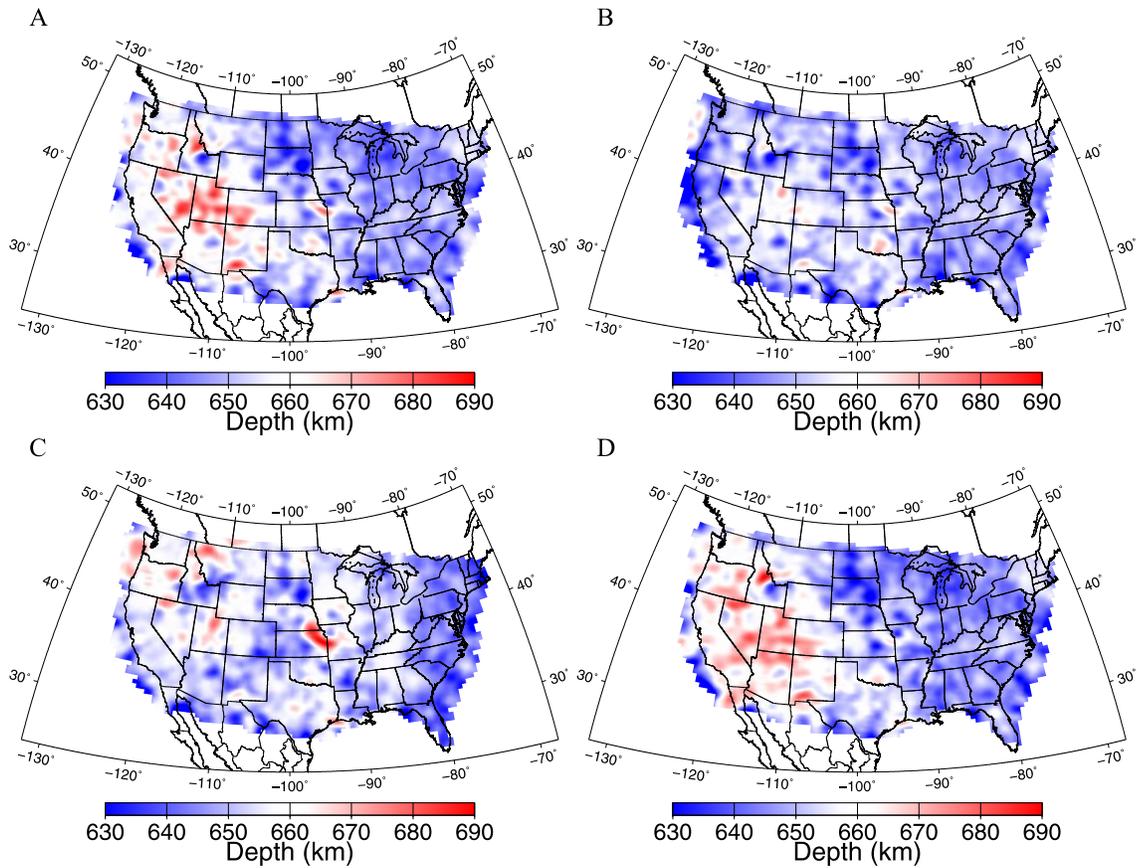

Figure 3. Topography of d660 corrected from four tomography models. The tomography models used in A, B, C, and D are MIT15, US16, USSL14, and GyPSuM correspondingly. The black lines are state boundaries.

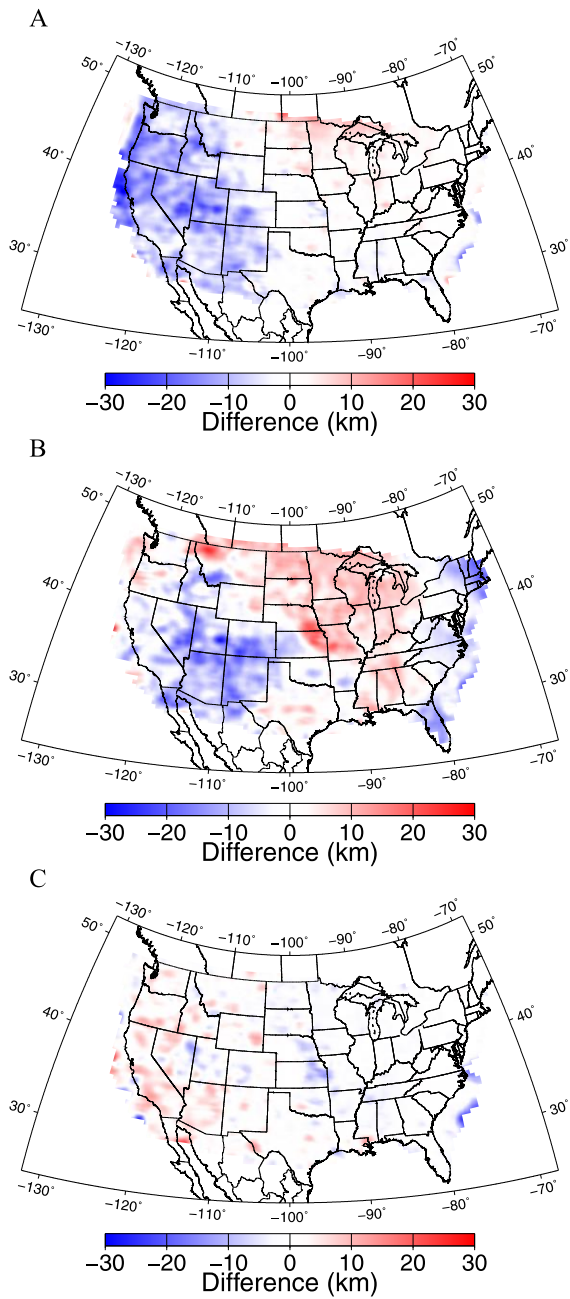

Figure 4. The difference of d660's topography relative to MIT15 reference. These are computed as d660 topography from other models subtract that from MIT15. A is from US16 model. B is from USSL14 model. C is from GyPSuM model.

Comparing the difference of d410 for each model with that of d660, we can see that correction for the d410 and d660 follows a similar large-scale trend in each of the tomography models. The difference is around the horizontal scale of 100 km or lower, and mostly within 10 km range.

## 4 Discussion
### 4.1 Large-scale roughness of the discontinuities

In our previous paper (Wang & Pavlis, 2016) we noted the large-scale offset of d410 and d660 between the western and eastern US. Gao & Liu (2014) had previously found the same feature with CCP stacking. We suggested in our earlier paper that we were concerned that the observed offset was a migration artifact resulting from an inadequate migration velocity. Our results here confirm that suggestion was probably correct and the 50 km scale depression of d410 and d660 in the western US is likely a velocity model artifact. Both MIT15 and GyPSuM show this large-scale feature, but it is much lower amplitude in US16 and USSL14. We briefly discuss why we think this suggest US16 and USSL14 are closer to reality.

The strongest evidence that the apparent offset in d410 and d660 between the eastern and western US is a velocity artifact is seen in Figure 5. Figure 5 shows that all four models produce a nearly identical isopach map for transition zone thickness. That result is an evidence that the velocity structure that produces the apparent offset is above d410, but the models are not in agreement on the form and size of velocity variation. To further assess what is and is not real we need to review some details of these four models.

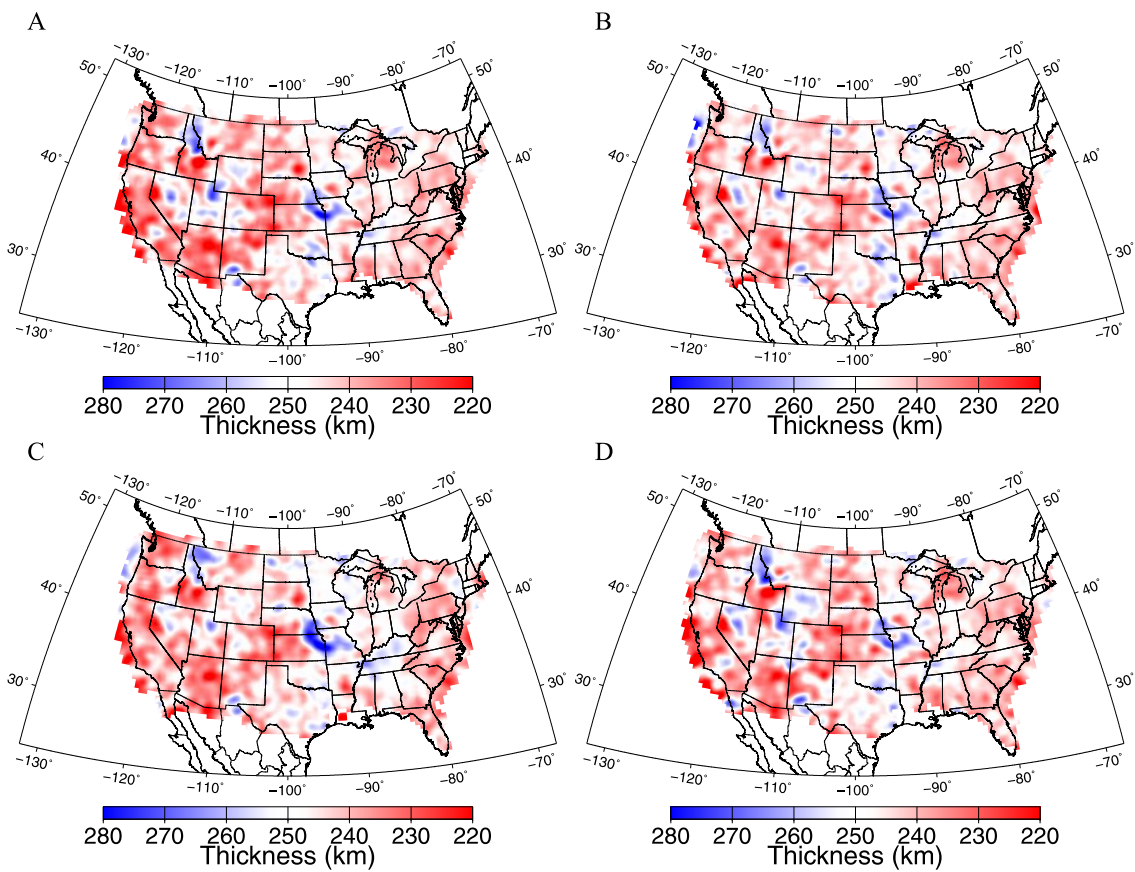

Figure 5. Thickness of the mantle transition zone from images migrated with four tomography models. The tomography models used in A, B, C, and D are MIT15, US16, USSL14, and GyPSuM respectively.

The US16 model is mainly constrained by surface wave dispersion data. It only has velocity estimates to 150 km depth. Furthermore, it is well known that surface wave inversion resolution decreases progressively with depth to the point that at the base of the model the resolution is very poor. A more important point for this paper is that the model contains larger velocity variations than either MIT15 or GyPSuM. For that reason, the surface generated from US16 elevates the western part of both d410 and d660 significantly. US16 seems to be doing a better job of capturing the size of the very low wave speed in the upper mantle centered on the Basin and Range.

The USSL14 model is a body wave tomography model that starts at 60 km and has velocity perturbations estimated to a depth of 1220 km. This model elevates d410 and d660 in the southwest US and depressing them in the northeast US In addition to correcting the large-scale general trend it also produces significant variation from the other models at an intermediate scale of several hundred kilometers. This is presumably a result of the intrinsic higher resolution the body wave tomography data added to a lower resolution 3D starting model in their inversion. Our results suggest a similar enhancement of the results with USSL14 within the transition zone. Figure 5 shows that the thickness maps for the other models are nearly identical, but USSL14 shows significant differences at intermediate scales. This almost certainly reflects stronger velocity perturbations in the transition zone for USSL14 than the other models. The key feature of USSL14 that causes it to create larger variations in d410 and d660 topography is that is uses a surface wave model with larger upper mantle variations as a starting model for a body wave inversion. That approach yields an overall perturbation that is much larger than MIT15 which is based on only P wave arrival times.

The GyPSuM is also a body wave tomography model, but it was computed by inverting travel times at the global scale. Thus, it is highly smoothed. Of more importance to this paper, from the surface to the core-mantle boundary perturbations defined by GyPSuM are an order of magnitude smaller than those found in the USSL14 model. These relatively small perturbations lead to much smaller corrections on both discontinuities' topography. MIT15 shares this deficiency in perturbation amplitude. Pavlis et al. (2012) compared all tomography models estimated from USArray data and noted the version of this model available then (MIT15) has much smaller perturbations than any of the other models they examined. We found the d410 and d660 surfaces produced from GyPSum and MIT15 were largely identical with any differences attributable to the human error from manual-picking the horizons. That also explains the increased topography difference for d660, which has greater error from decreased vertical resolution and signal-to-noise ratio at that depth. For the same reason, we consider the topography difference that are small in amplitude and horizontal scale from all other tomography models to be the same type of error. Such an error is extremely difficult to avoid especially when interpreting and comparing many different results. We tried to minimize its effect by having the same person pick these horizons in a short time frame, since a specially designed auto-tracker is yet to made for this type of analysis.

### 4.2 Static correction for body wave tomography model

Given that isopach maps show all models we examined produce a nearly constant thickness transition zone, a relevant question is how much of the observed topography can be attributed to variations in crustal thickness alone? Variations in crustal thickness are completely analogous to static corrections in seismic reflection data with the crust acting like the weathered layer. To test the scale of this effect, we implement a static correction using the CRUST1.0 model. We compute a static at each station as the S-P travel time difference for vertical incidence between the CRUST1.0 model and AK135 reference model. The static corrected receiver functions are then migrated as before. We show results here from USSL14, but the effects are nearly identical for all models.

The topography generated from this result shows that the effects of static corrections alone are small compared to apparent topography created by upper mantle structure (Figure 5A, B). Figure 5C and 5D show that effects of the static corrections are all less than 5 km and, as expected for a static correction, are nearly identical for d410 and d660. Also, the overall result with statics applied elevates both discontinuities. This is, no doubt, a reflection of AK135 having a different average structure from that tabulated for CRUST1.0 in the US Since the corrections are very close to the error from interpretation, a crustal static correction based on CRUST1.0, at least, is relatively small compared to the effects of upper mantle structure inferred from tomography models.

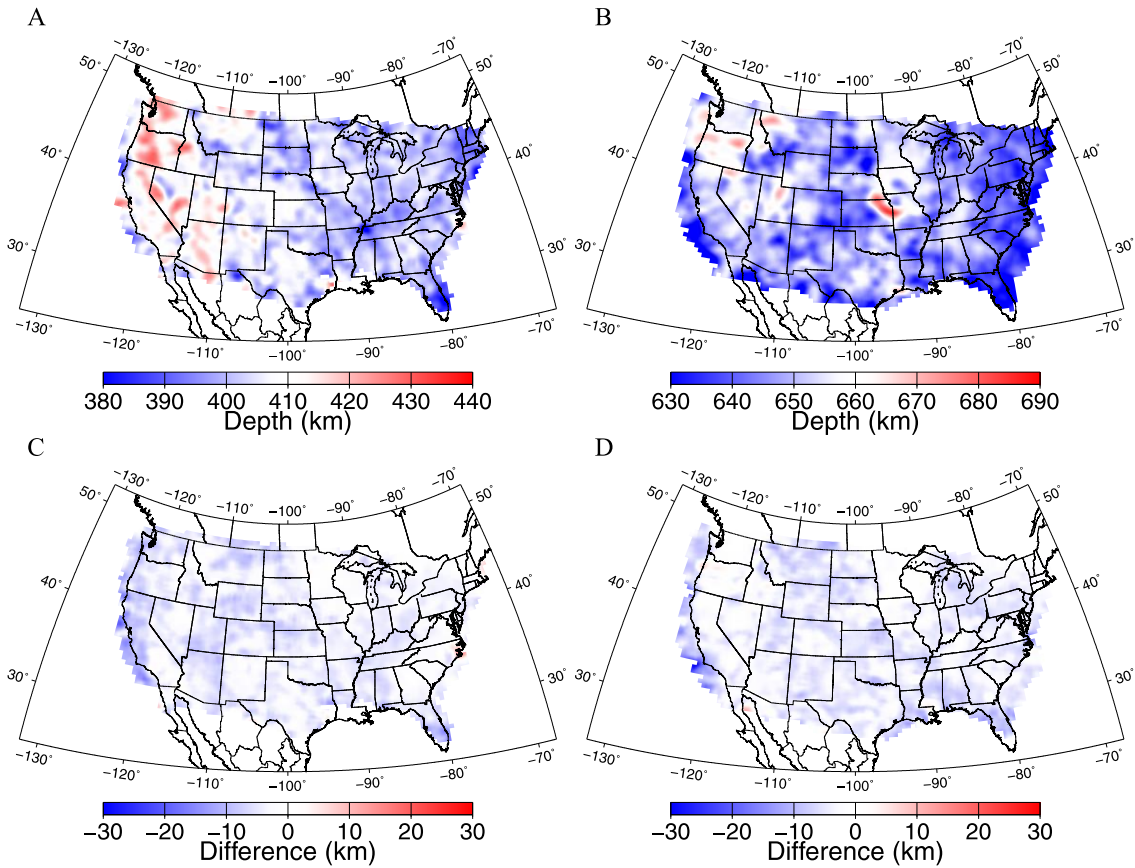

Figure 6. Static correction results for d410 and d660. These results are produced from receiver functions corrected by the CRUST1.0 model and migrated with the USSL14 model. A and B are the topography of d410 and d660 accordingly. C and D are the difference from the comparable surface without static correction shown in Figure 1C and Figure 3C.

### 4.3 Small-scale roughness of the discontinuities

A significant feature that persists across results from different tomography models is the topography variation at the scale of the order of 100 km. This is the major observation in Wang & Pavlis (2016) and our results confirm this observation is not dependent on the choice of tomography models. Although the regional body wave tomography model USSL14 does have some topography corrections near this scale, they are not as strong and pervasive for both d410 and d660 and do not cancel out or reduce the small-scale roughness features on the topography. Some of the topography corrections in these results are even smaller than 100 km, but we assert that they are from interpretation errors based on their random pattern in the results from the GyPSuM model. We also found the isopach from different tomography models to be almost identical (Figure 5). It implies that despite the presence of velocity perturbation within the transition zone for the USSL14 and GyPSuM model, none of them are significant enough to interfere with the imaged topography on d660. Moreover, a lot of small-scale roughness features are observed from these thickness maps — they are not cancelled out from subtracting d410 and d660. This suggests strong that small scale roughness is unlikely an artifact from upper mantle structures not captured with current generation of tomography models.

## 4.4 Implication on the Mantle Dynamics

We further separate the small-scale features from large-scale features by applying a low-pass spatial filter with 1,000 km diameter to the isopach map from the USSL14 model (Figure 7). Figure 7 shows that the transition zone in the southwestern US is thinner overall than the rest of lower 48 states. The standard model for the discontinuities as phase transitions would imply the transition zone in that area has a higher temperature than elsewhere in the US. We suggest that result is likely a direct reflection of mantle upwelling in that region created by the Farallon slab window. Figure 7 illustrates a model for the northern edge of the slab window from Pavlis et al. (2012). It also shows a comparable, less commonly appreciated slab window edge to the south reconstructed by Panessa (2013). Panessa (2013) reconstructed this edge by assuming a continuous subduction zone existed in western North America until 29 Ma when the slab window initiated. He then merged two plate motion models to define the transition from a continuous subduction zone prior to 29 Ma to the present situation with independent motion of the Cocos Plate in Central America from the remnant of the Farallon Plate now present in the Pacific Northwest. In particular, Panessa (2013) used the Doubrovine & Tarduno (2008) Farallon-North America for stage poles prior to 20 MA and those given by Pindell et al. (1988) for the Cocos Plate after 20 Ma. Panessa's edge model has complexities related to the opening of the Gulf of California and a related southward jump of the subduction system around 12.5 Ma. That ambiguity makes the north-south position of the southern slab window edge shown in Figure 7 uncertain, but the flow line direction that defines the strike of the southern edge is well defined by plate kinematics.

The area where the transition zone is thinner, and thus likely higher temperature, correlates remarkably well with the edges of the slab window predicted with the rigid plate motion models used by Pavlis et al. (2012) and Panessa (2013). Pavlis et al. (2012) also noted that high velocity bodies in the upper mantle in the southwestern US linked to the slab window edge were one of the most robust features seen in the suite of tomography models they examined. That suggests strongly that mantle upwelling is occurring in that area linked to the slab window. How the related sinking motion related to the Farallon slab north and south of the slab window is less clear. Our results suggest the southern edge may not differ greatly from flow lines predicted from plate motion because the edge of the thinned transition zone under Texas is very close in strike to that of the slab window edge predicted from backtracking Cocos Plate motion. The northern edge shows more variation. This likely reflects the more complex Cenozoic history of that region related to the subduction of the Kula plate, accretion of the Siletzia terrane (e.g. Schmandt & Humphreys, 2011), the Yellowstone volcanic system, and/or the possible idea that the Cenozoic slab has disaggregated within the mantle under North America (e.g. Liu & Stegman, 2011).

In the central US Figure 7A shows a thickening of the transition zone from Minnesota through a strong maximum centered on the Missouri-Kansas border. We note that this marks the region above where most tomography models define the Farallon Slab in the lower mantle (e.g. Sigloch, 2011).

That association suggests this is a region cooled by downward motion driven by the old Farallon slab under eastern North America.

The high-pass filtered isopach map shows equally intriguing features. The small-scale roughness has a higher amplitude in the western US We also suggest these small-scale irregularities have a fabric oriented roughly perpendicular to plate motion. Few would disagree that the western U.S mantle is more likely to be characterized by larger vertical flow no matter what model one builds to define the overall flow process. It is also relatively stronger near the Cascadia subduction and Laramide slab, where the subducted slab is observed at transition zone depth (Sigloch, 2011). Wang & Pavlis (2016) noted this correlation from the discontinuity surfaces alone, but Figure 7 strengthens that observation. These new observations provide additional support for our interpretation that the small-scale roughness relates to the vertical mantle flow through the discontinuities influenced by the large-scale mantle dynamics.

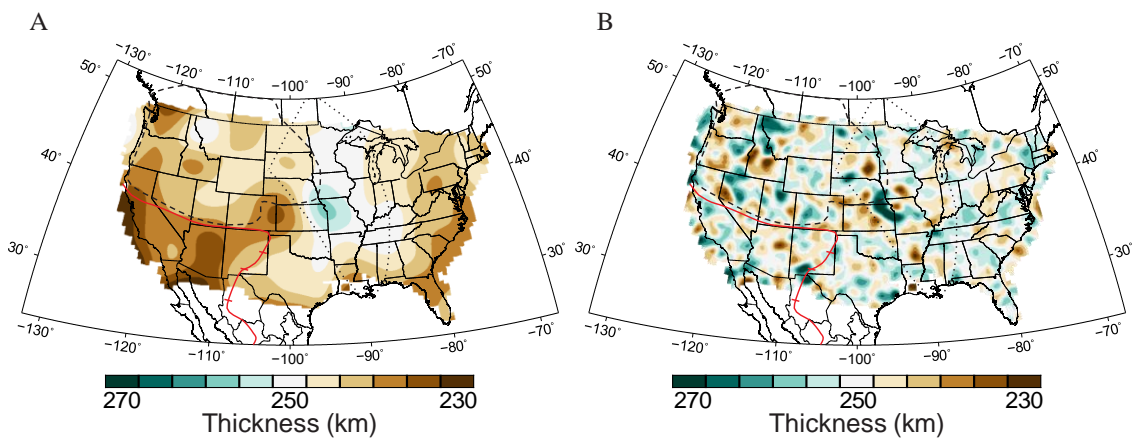

Figure 7. Filtered thickness map of the mantle transition zone from images migrated with USSL14 model. A is low-pass filtered with Gaussian function of 1,000 km width. B is the result subtracting A from the original isopach map (Figure 5C), which is equivalent to high-pass filtering. The solid red line is the edge of the Farallon slab window predicted by Panessa (2013). The marks on top of the southern edge (red curve) mark the subduction ages of 15, 20, 25 Ma from southwest to northeast. The areas circled with dashed line and dotted line are areas of Cascadia subduction and Laramide slab respectively observed by Sigloch (2011).

## 5 Conclusion

We present new imaging results of the d410 and d660 using receiver functions migrated by PWMIG method with a selection of tomography models. Both surface wave and body wave tomography models can reduce the large-scale topography variations, which is mainly affected by complexities in the upper mantle, but pure body wave models consistently underestimate such large-scale velocity difference. Combining corrections from body wave tomography model and crustal static correction does not further reduce the large-scale feature. In contrast, the small-scale roughness feature on both d410 and d660 stays constant across different tomography models. We assert such roughness is not likely produced from artifact from uncaptured upper mantle structures. The new results reveal the connection between the small-scale roughness with large-scale mantle provinces suggesting such roughness as an indicator of large-scale vertical flows across the transition zone.

**Acknowledgement**

We acknowledge Schlumberger for providing access to the Petrel seismic interpretation software package. Data from the TA network were made freely available as part of the EarthScope USArray facility, operated by Incorporated Research Institutions for Seismology (IRIS) and supported by the National Science Foundation, under Cooperative Agreements EAR-1261681. The cyberinfrastructure of this research was supported in part by Lilly Endowment, Inc., through its support for the Indiana University Pervasive Technology Institute, and in part by the Indiana METACyt Initiative. The Indiana METACyt Initiative at IU is also supported in part by Lilly Endowment, Inc. Support for this work came from the US National Science Foundation Earthscope Program under award EAR-1358149.

**References**

Ai, Y., Chen, Q., Zeng, F., Hong, X., & Ye, W. (2007). The crust and upper mantle structure beneath southeastern China. *Earth and Planetary Science Letters*, *260*(3–4), 549–563. https://doi.org/10.1016/j.epsl.2007.06.009

Benz, H. M., & Vidale, J. E. (1993). Sharpness of upper-mantle discontinuities determined from high-frequency reflections. *Nature*, *365*(6442), 147–150. https://doi.org/10.1038/365147a0

Bernal, J. D. (1936). Geophysical discussion. *Observatory*, *59*, 265–269.

Burdick, S., Vernon, F. L., Martynov, V., Eakins, J., Cox, T., Tytell, J., et al. (2017). Model Update May 2016: Upper-Mantle Heterogeneity beneath North America from Travel-Time Tomography with Global and USArray Data. *Seismological Research Letters*, *88*(2A), 319–325. https://doi.org/10.1785/0220160186

Cao, A., & Levander, A. (2010). High-resolution transition zone structures of the Gorda Slab beneath the western United States: Implication for deep water subduction. *Journal of Geophysical Research: Solid Earth*, *115*(B7), B07301. https://doi.org/10.1029/2009JB006876

Chevrot, S., Vinnik, L., & Montagner, J.-P. (1999). Global-scale analysis of the mantle Pds phases. *Journal of Geophysical Research: Solid Earth*, *104*(B9), 20203–20219. https://doi.org/10.1029/1999JB900087

Collier, J. D., & Helffrich, G. R. (1997). Topography of the "410" and "660" km seismic discontinuities in the Izu-Bonin Subduction Zone. *Geophysical Research Letters*, *24*(12), 1535–1538. https://doi.org/10.1029/97GL01383

Cottaar, S., & Deuss, A. (2016). Large-scale mantle discontinuity topography beneath Europe: Signature of akimotoite in subducting slabs. *Journal of Geophysical Research: Solid Earth*, 2015JB012452. https://doi.org/10.1002/2015JB012452

Doubrovine, P. V., & Tarduno, J. A. (2008). A revised kinematic model for the relative motion between Pacific oceanic plates and North America since the Late Cretaceous. *Journal of Geophysical Research: Solid Earth*, *113*(B12), B12101. https://doi.org/10.1029/2008JB005585

Dueker, K. G., & Sheehan, A. F. (1997). Mantle discontinuity structure from midpoint stacks of converted P to S waves across the Yellowstone hotspot track. *Journal of Geophysical Research: Solid Earth*, *102*(B4), 8313–8327. https://doi.org/10.1029/96JB03857

Engdahl, E. R., & Flinn, E. A. (1969). Seismic Waves Reflected from Discontinuities within Earth's Upper Mantle. *Science*, *163*(3863), 177–179. Retrieved from http://www.jstor.org/stable/1725580

Fan, C., Pavlis, G. L., & Tuncay, K. (2006). GCLgrid: A three-dimensional geographical curvilinear grid library for computational seismology. *Computers & Geosciences*, *32*(3), 371–381. https://doi.org/10.1016/j.cageo.2005.07.001

Flanagan, M. P., & Shearer, P. M. (1998). Global mapping of topography on transition zone velocity discontinuities by stacking SS precursors. *Journal of Geophysical Research: Solid Earth*, *103*(B2), 2673–2692. https://doi.org/10.1029/97JB03212

Gao, S. S., & Liu, K. H. (2014). Mantle transition zone discontinuities beneath the contiguous United States. *Journal of Geophysical Research: Solid Earth*, *119*(8), 6452–6468. https://doi.org/10.1002/2014JB011253

Gu, Y., Dziewonski, A. M., & Agee, C. B. (1998). Global de-correlation of the topography of transition zone discontinuities. *Earth and Planetary Science Letters*, *157*(1–2), 57–67. https://doi.org/10.1016/S0012-821X(98)00027-2

Helffrich, G. (2000). Topography of the transition zone seismic discontinuities. *Reviews of Geophysics*, *38*(1), 141. https://doi.org/10.1029/1999RG000060

Jeffreys, H. (1952). *The Earth, Its Origin, History, and Physical Constitution*. Cambridge University Press.

Kennett, B. L. N., Engdahl, E. R., & Buland, R. (1995). Constraints on seismic velocities in the Earth from traveltimes. *Geophysical Journal International*, *122*(1), 108–124. https://doi.org/10.1111/j.1365-246X.1995.tb03540.x

Laske, G., Masters, G., Ma, Z., & Pasyanos, M. (2013). Update on CRUST1.0 - A 1-degree Global Model of Earth's Crust (Vol. 15, pp. EGU2013-2658). Presented at the EGU General Assembly Conference Abstracts. Retrieved from http://adsabs.harvard.edu/abs/2013EGUGA..15.2658L

Liu, L., & Stegman, D. R. (2011). Segmentation of the Farallon slab. *Earth and Planetary Science Letters*, *311*(1–2), 1–10. https://doi.org/10.1016/j.epsl.2011.09.027

Melbourne, T., & Helmberger, D. (1998). Fine structure of the 410-km discontinuity. *Journal of Geophysical Research: Solid Earth*, *103*(B5), 10091–10102. https://doi.org/10.1029/98JB00164

Morozova, E. A., Morozov, I. B., Smithson, S. B., & Solodilov, L. N. (1999). Heterogeneity of the uppermost mantle beneath Russian Eurasia from the ultra-long-range profile quartz. *Journal of Geophysical Research: Solid Earth*, *104*(B9), 20329–20348. https://doi.org/10.1029/1999JB900142

Neele, F. (1996). Sharp 400-km discontinuity from short-period P reflections. *Geophysical Research Letters*, *23*(5), 419–422. https://doi.org/10.1029/96GL00378

Panessa, A. (2013). *Evidence for a coherent southeastern edge of the Farallon slab window from USArray tomography models* (M.S.). Retrieved from http://search.proquest.com/docview/1318856728/abstract/155B4374E920424CPQ/1

Pavlis, G. L. (2011). Three-dimensional, wavefield imaging of broadband seismic array data. *Computers & Geosciences*, *37*(8), 1054–1066. Retrieved from http://www.sciencedirect.com/science/article/pii/S0098300411000185

Pavlis, G. L., Sigloch, K., Burdick, S., Fouch, M. J., & Vernon, F. L. (2012). Unraveling the geometry of the Farallon plate: Synthesis of three-dimensional imaging results from USArray. *Tectonophysics*. Retrieved from http://www.sciencedirect.com/science/article/pii/S0040195112000856

Pindell, J. L., Cande, S. ., Pitman, W. C., Rowley, D. B., Dewey, J. F., Labrecque, J., & Haxby, W. (1988). A plate-kinematic framework for models of Caribbean evolution. *Tectonophysics*, *155*(1–4), 121–138. https://doi.org/10.1016/0040-1951(88)90262-4

Poppeliers, C., & Pavlis, G. L. (2003a). Three-dimensional, prestack, plane wave migration of teleseismic *P* -to- *S* converted phases: 1. Theory. *Journal of Geophysical Research*, *108*(B2). https://doi.org/10.1029/2001JB000216

Poppeliers, C., & Pavlis, G. L. (2003b). Three-dimensional, prestack, plane wave migration of teleseismic *P* -to- *S* converted phases: 2. Stacking multiple events. *Journal of Geophysical Research*, *108*(B5). https://doi.org/10.1029/2001JB001583

Priestley, K., Cipar, J., Egorkin, A., & Pavlenkova, N. (1994). Upper-mantle velocity structure beneath the Siberian platform. *Geophysical Journal International*, *118*(2), 369–378. https://doi.org/10.1111/j.1365-246X.1994.tb03968.x

Revenaugh, J., & Jordan, T. H. (1989). A study of mantle layering beneath the western Pacific. *Journal of Geophysical Research: Solid Earth*, *94*(B5), 5787–5813. https://doi.org/10.1029/JB094iB05p05787

Revenaugh, J., & Jordan, T. H. (1991). Mantle layering from ScS reverberations: 2. The transition zone. *Journal of Geophysical Research: Solid Earth*, *96*(B12), 19763–19780. https://doi.org/10.1029/91JB01486

Ringwood, A. E. (1956). The Olivine–Spinel Transition in the Earth's Mantle. *Nature*, *178*(4545), 1303–1304. https://doi.org/10.1038/1781303a0

Schmandt, B., & Humphreys, E. (2011). Seismically imaged relict slab from the 55 Ma Siletzia accretion to the northwest United States. *Geology*, *39*(2), 175–178. https://doi.org/10.1130/G31558.1

Schmandt, B., & Lin, F.-C. (2014). *P* and *S* wave tomography of the mantle beneath the United States. *Geophysical Research Letters*, *41*(18), 6342–6349. https://doi.org/10.1002/2014GL061231

Shearer, P. M. (1991). Constraints on upper mantle discontinuities from observations of long-period reflected and converted phases. *Journal of Geophysical Research: Solid Earth*, *96*(B11), 18147–18182. https://doi.org/10.1029/91JB01592

Shearer, P. M. (1993). Global mapping of upper mantle reflectors from long-period SS precursors. *Geophysical Journal International*, *115*(3), 878–904. https://doi.org/10.1111/j.1365-246X.1993.tb01499.x

Shen, W., & Ritzwoller, M. H. (2016). Crustal and uppermost mantle structure beneath the United States. *Journal of Geophysical Research: Solid Earth*, *121*(6), 2016JB012887. https://doi.org/10.1002/2016JB012887

Sigloch, K. (2011). Mantle provinces under North America from multifrequency P wave tomography. *Geochemistry, Geophysics, Geosystems*, *12*(2), Q02W08. https://doi.org/10.1029/2010GC003421

Simmons, N. A., Forte, A. M., Boschi, L., & Grand, S. P. (2010). GyPSuM: A joint tomographic model of mantle density and seismic wave speeds. *Journal of Geophysical Research: Solid Earth*, *115*(B12), B12310. https://doi.org/10.1029/2010JB007631

Thybo, H., Nielsen, L., & Perchuc, E. (2003). Seismic scattering at the top of the mantle Transition Zone. *Earth and Planetary Science Letters*, *216*(3), 259–269. https://doi.org/10.1016/S0012-821X(03)00485-0

Tibi, R., & Wiens, D. A. (2005). Detailed structure and sharpness of upper mantle discontinuities in the Tonga subduction zone from regional broadband arrays. *Journal of Geophysical Research: Solid Earth*, *110*(B6), B06313. https://doi.org/10.1029/2004JB003433

Vidale, J. E., Ding, X.-Y., & Grand, S. P. (1995). The 410-km-depth discontinuity: A sharpness estimate from near-critical reflections. *Geophysical Research Letters*, *22*(19), 2557–2560. https://doi.org/10.1029/95GL02663

Wang, Y., & Pavlis, G. L. (2016). Roughness of the mantle transition zone discontinuities revealed by high-resolution wavefield imaging. *Journal of Geophysical Research: Solid Earth*, *121*(9), 6757–6778. https://doi.org/10.1002/2016JB013205

Yamazaki, A., & Hirahara, K. (1994). the thickness of upper mantle discontinuities, as inferred from short-period J-array data. *Geophysical Research Letters*, *21*(17), 1811–1814. https://doi.org/10.1029/94GL01418

Yang, X. (2016). *Seismic imaging of the lithosphere beneath the southern Illinois Basin and its tectonic implications* (Ph.D.). Retrieved from https://search.proquest.com/docview/1848960794/abstract/E3B74EF7DED84686PQ/1